\documentclass[a4paper]{PoS}
\usepackage{mathrsfs}
\usepackage{lineno}
\usepackage{amsmath}

\title{Searching for Dark Matter decay signals in the Galactic Halo with the MAGIC telescopes}

\ShortTitle{Dark Matter decay in the Galactic Halo with MAGIC}

\author{\speaker{D. Ninci}\\
        Institut de Fisica d'Altes Energies (E-08193 Barcelona, Spain)\\
        E-mail: \email{dninci@ifae.es}}
\author{T. Inada$^f$, J. Rico$^a$, D. Kerszberg$^a$, M. Doro$^b$, M. Vazquez Acosta$^c$, S. Lombardi$^d$, C. Maggio$^e$, M. H\"utten$^g$ on behalf of the MAGIC Collaboration\footnote{\texttt{https://magic.mpp.mpg.de/}. For collaboration list see PoS(ICRC2019)1177}, \\
\llap{$^a$}Institut de Fisica d'Altes Energies (E-08193 Barcelona, Spain),\\
\llap{$^b$}Universit\`a di Padova and INFN, I-35131 Padova, Italy,\\
\llap{$^c$}Inst. de Astrof\'isica de Canarias, E-38200 La Laguna, and Universidad de La Laguna, Dpto. Astrof\'isica, E-38206 La Laguna, Tenerife, Spain, \\
\llap{$^d$}National Institute for Astrophysics (INAF), I-00136 Rome, Italy, \\
\llap{$^e$}Departament de F\'isica, and CerES-IEEC, Universitat Aut\`onoma de Barcelona, E-08193 Bellaterra, Spain, \\
\llap{$^f$}Japanese MAGIC Consortium: ICRR, The University of Tokyo, 277-8582 Chiba, Japan; Department of Physics, Kyoto University, 606-8502 Kyoto, Japan; Tokai University, 259-1292 Kanagawa, Japan; RIKEN, 351-0198 Saitama, Japan \\
\llap{$^g$}Max-Planck-Institut f\"ur Physik, D-80805 M\"unchen, Germany \\

}

\abstract{MAGIC is a system of two Cherenkov telescopes located in the Canary island of La Palma. A key part of MAGIC Fundamental Physics program is the search for indirect signals of Dark Matter (DM) from different sources. In the Milky Way, DM forms an almost spherically symmetric halo, with a density peaked towards the center of the Galaxy and decreasing toward the outer region. We search for DM decay signals from the Galactic Halo, with a special methodology developed for this work. Our strategy is to compare pairs of observations performed at different angular distances from the Galactic Center, selected in such a way that all the diffuse components cancel out, except for those coming from the DM. In order to keep the systematic uncertainty of this novel background estimation method down to a minimum, the observation pairs have been acquired during the same nights and follow exactly the same azimuth and zenith paths. We collected 20 hours of data during 2018. Using half of them to determine the systematic uncertainty in the background estimation of our analysis, we obtain a value of 4.8\% with no dependence on energy. Accounting for this systematic uncertainty in the likelihood analysis based on the 10 remaining hours of data collected so far, we present the limit to TeV DM particle with a lifetime of $10^{26}$ s in the $\mathrm{b\bar{b}}$ decay channel. }

\FullConference{36th International Cosmic Ray Conference -ICRC2019-\\
		July 24th - August 1st, 2019\\
		Madison, WI, U.S.A.}

\begin{document}

\section{Introduction}
The current Cosmological Model affirms that about 85\% of the Universe is composed by an unknown form of matter, called Dark Matter (DM), likely consisting of undetected relic particles from the Big Bang. The evidence supporting its existence arises from
a variety of astrophysical and cosmological observations.
Many well-motivated DM particle candidates have been proposed in scenarios of physics beyond the Standard Model~\cite{General}.  A generic class of a DM particle candidate is the Weakly Interacting Massive Particle (WIMP) which could annihilate or decay into Standard Model particles with rates resulting in fluxes at Earth detectable with current instruments. 
Albeit the annihilation case is more discussed in the literature, the search for signals of DM decay
is also well motivated, provided that the lifetime of DM particle is larger than age of the Universe (see e.g.~\cite{dmmilkyway}).
 	

Regardless of its specific nature, the DM is supposed to form an almost spherical halo around galaxies such as the Milky Way. The DM density distribution is peaked toward the center of the Galaxy and decreases towards the outer region~\cite{DMHalo}. The expected $\mathrm{\gamma}$-ray flux from DM from a region $\mathrm{\Delta \Omega}$ centred at an angular distance $\varphi$ from the Galactic Center (GC) can be expressed with the formula:
$$
\frac{d\phi}{dE}(\varphi,\, \Delta  \Omega) =
\frac{\alpha} {4\pi\cdot m_{DM}}\frac{dN}{dE}\cdot J(\varphi, \Delta \Omega) \qquad \mathrm{where} \qquad 
J(\varphi, \Delta \Omega) = \int_{\Delta \Omega} d\Omega \int_{l.o.s} dl \rho^{\beta}(l(\varphi))
$$ 
where $m_{DM}$ is the DM particle mass, $\frac{dN}{dE}$ is the average photon spectrum after an annihilation or decay process. J is the astrophysical factor as defined above (also called the J-factor), the integrals to compute J run over the line of sight (l.o.s.) defined by $\varphi$ and the total field of view (FoV) and depend on $\rho$ the DM density. For annihilation $\alpha=\frac{<\sigma v>}{2m_{DM}}$ and $\beta=2$, while for decay $\alpha=\frac{1}{\tau_{DM}}$ and $\beta=1$, where $<\sigma v>$ is the thermal average annihilation cross section and $\tau_{DM}$ is the DM lifetime.


We present the results from observations of the Galactic Halo (GH) surrounding the Galaxy, searching for DM decay signals with the MAGIC telescopes.

\section{MAGIC observations of the Galactic Halo}
MAGIC is a system of two 17 meter Cherenkov telescopes, located at the Roque de Los Muchachos observatory on the Canary island of La Palma. They detect the Cherenkov light created by the particle showers initiated by cosmic-rays and $\gamma$-rays entering the Earth atmosphere. MAGIC is operative since 2004 as a single telescope and since 2009 with two telescopes.

Albeit the GC benefits of a larger J-factor with respect to the GH~\cite{HESS2}, it is a very crowded region with a very large astrophysical background which is highly model dependent. 
Moreover, MAGIC can only observe the GC at a large zenith distance (Zd), implying higher energy threshold and lower sensitivity to DM processes with respect to the small Zd observations.
However, given the extension of the GH, MAGIC can observe regions for which $\varphi>10^\circ$ at low Zd, thus increasing the sensitivity while avoiding the most crowded region. 

\begin{figure}[t]
\centering
\includegraphics[width=0.50\textwidth]{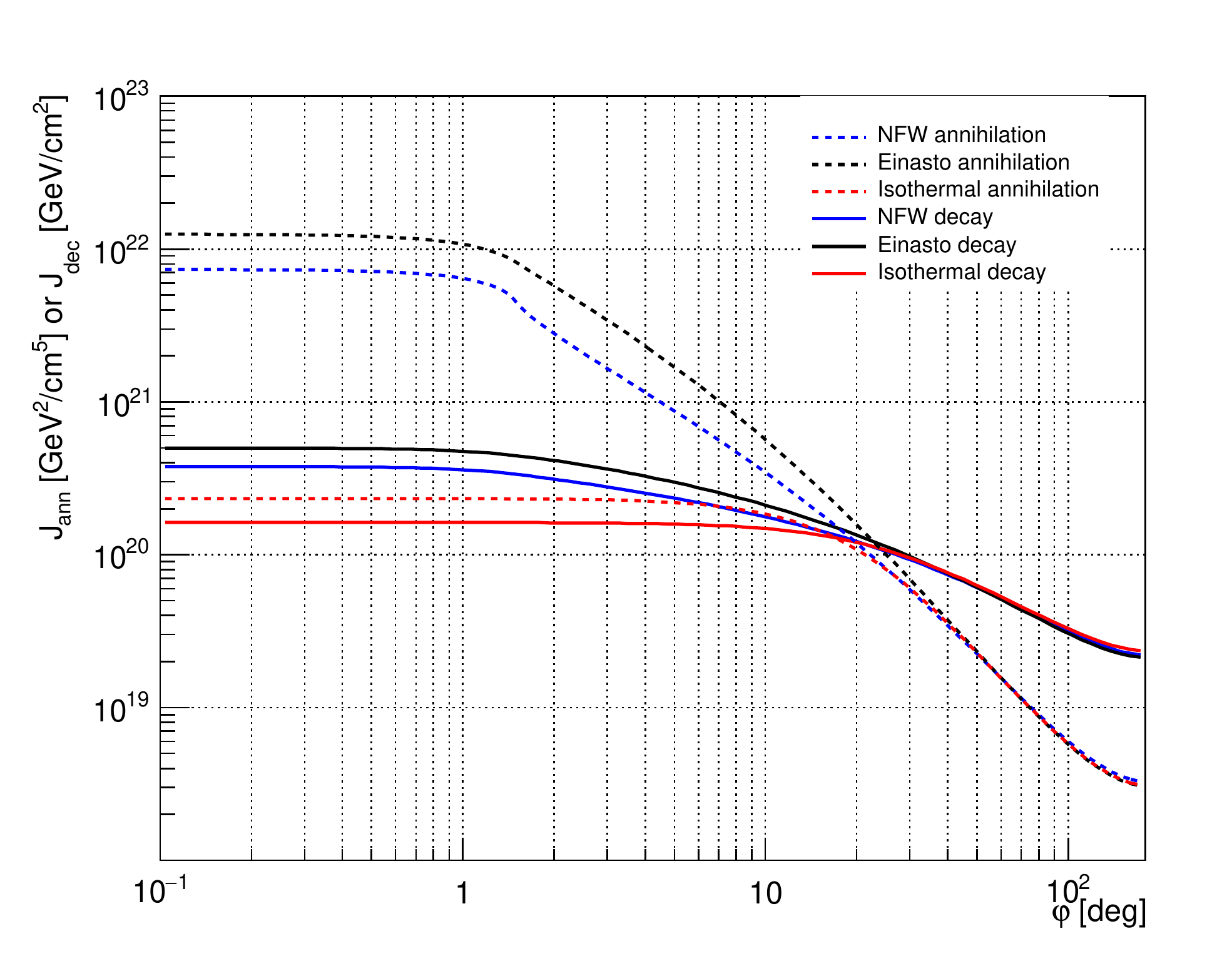}
\caption{J-factor as a function of $\varphi$ for DM annihilation and decay processes. The J-factor is computed for a $\Delta\Omega$ with an angular radius of $1.5^\circ$.  The curves are computed for three DM density models of the Milky Way: the cuspy Einasto and Navarro-Frenk-White (NFW) profiles, and an isothermal (cored) profile~\cite{Ibarra2010}. For the decay case, the three profiles are almost identical inside the region of interest for this work ($\varphi>10^\circ$).}
\label{fig:myjfactor}
\end{figure}

Moving out through the GH ($\varphi > 10^\circ$) the expected flux from DM annihilation drops significantly with respect to the GC observations ($\phi(\varphi=90^\circ)/\phi(\varphi=0^\circ)_{ann}\approx10^{-4}$, see Fig.~\ref{fig:myjfactor}). This happens because the J-factor for annihilation depends quadratically on the DM density. Instead, since the J-factor for decay processes depends linearly on the DM density, the expected $\gamma$-ray flux from DM decay in the GH results to be comparable with the one expected from the GC (at most $\phi(\varphi = 90^\circ
)/\phi(\varphi = 0^\circ)]_{dec} \approx 10^{-1}$, see Fig.~\ref{fig:myjfactor}).

The expected DM signal from the GH is much more extended than the MAGIC FoV. Therefore we cannot use the standard observation mode and background estimation method that allow the simultaneous observation of the signal (ON) and background (OFF) regions. 
Consequently, we perform ON-OFF observations, with the ON observations performed
at relatively low $\varphi$ (i.e. with highest expected DM flux) while keeping low Zd, and the OFF at relatively high $\varphi$ (i.e. with lowest expected DM flux), in such a way that the difference of observed number of events is proportional to the DM flux. This can be demonstrated using the expected diffuse event rate $\mathcal{R}$ detected from a given observation, that can be written as:
$$
\mathcal{R} (l,\,b) = \mathcal{R}_{CR} + \mathcal{R}_{e^+e^-} + \mathcal{R}_{EG\mbox{-}\gamma}+\mathcal{R}_{EG\mbox{-}DM}+\mathcal{R}_{Gal\mbox{-}\gamma}(l,b) + \mathcal{R}_{DM}(l,\,b),
$$
where $(l, b)$ are the the Galactic longitude and latitude, $\mathcal{R}_{CR}$ is the event rate of cosmic rays,
$\mathcal{R}_{e^+ e^-}$ is the electron+positron event rate, $\mathcal{R}_{EG\mbox{-}\gamma}$ is the $\gamma$-ray event rate from integrated emission
of AGNs up to cosmological distances, $\mathcal{R}_{EG\mbox{-}DM}$ is the $\gamma$-ray event rate from extragalactic DM 
sources, $\mathcal{R}_{Gal\mbox{-}\gamma} (l, b)$ is the $\gamma$-ray event rate from the interactions of the cosmic rays with the
interstellar medium and from unresolved sources, and $\mathcal{R}_{DM} (l, b) $ is the $\gamma$-ray event rate from DM in the GH.
By subtracting $\mathcal{R}$ in observations of two different regions of the sky, all isotropic components cancel out: $$\mathcal{R}(l_1 , b_1 ) - \mathcal{R}(l_2 , b_2 ) = \mathcal{R}_{DM} (\varphi_1 ) - \mathcal{R}_{DM} (\varphi_2 ) + \mathcal{R}_{Gal\mbox{-}\gamma} (l_1 , b_1 ) - \mathcal{R}_{Gal\mbox{-}\gamma} (l_2 , b_2 ).$$
Since $\phi_{Gal\mbox{-}\gamma}$ is mildly anisotropic for $|b| > 10^\circ$ and for the energy accessible by MAGIC, we can select the FoVs for which $\Delta \mathcal{R}_{Gal-\gamma}$
is negligible, so that we obtain: $$\Delta \mathcal{R}(\varphi_1 , \varphi_2 ) = \mathcal{R}(l_1 , b_1 ) - \mathcal{R}(l_2 , b_2 ) = \mathcal{R}_{DM} (\varphi_1 ) - \mathcal{R}_{DM} (\varphi_2 ).$$ 
This expression only depends on $\varphi$, since only the DM decay contribution survives. $\mathcal{R}_{DM}$ is proportional to the DM flux, and therefore: $$\Delta \mathcal{R}(\varphi_1 , \varphi_2 ) = C(J(\varphi_1 ) - J(\varphi_2 )) = C\Delta J(\varphi_1 , \varphi_2 ),$$ where $C=\frac{1}{4\pi\tau m_{DM}}\cdot\frac{dN}{dE}$ and $\Delta J$ is the difference in the J-factor between the two pointing positions.

\begin{figure}[t]
\centering
\includegraphics[width=0.51\textwidth]{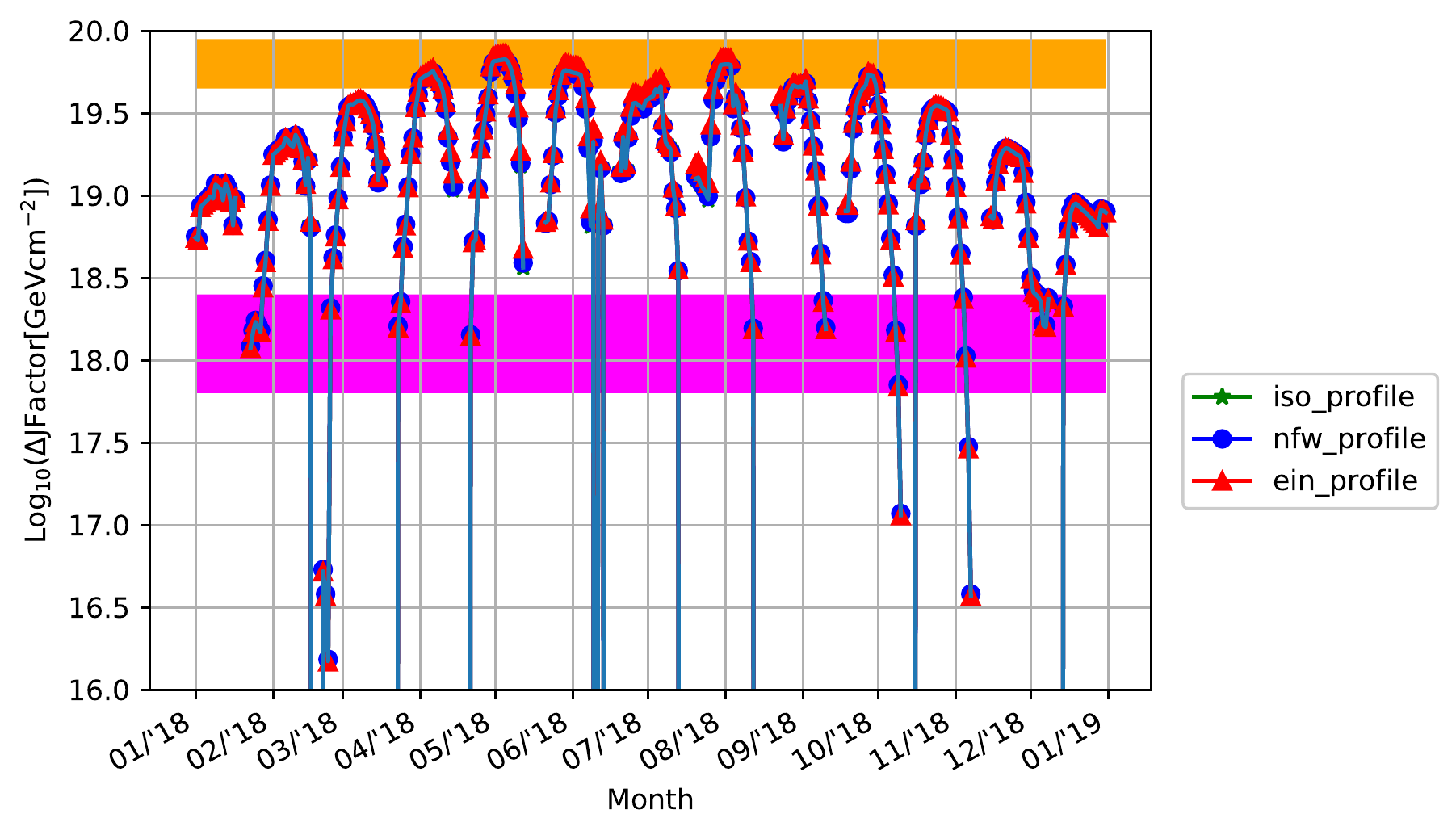}
\includegraphics[width=0.42\textwidth]{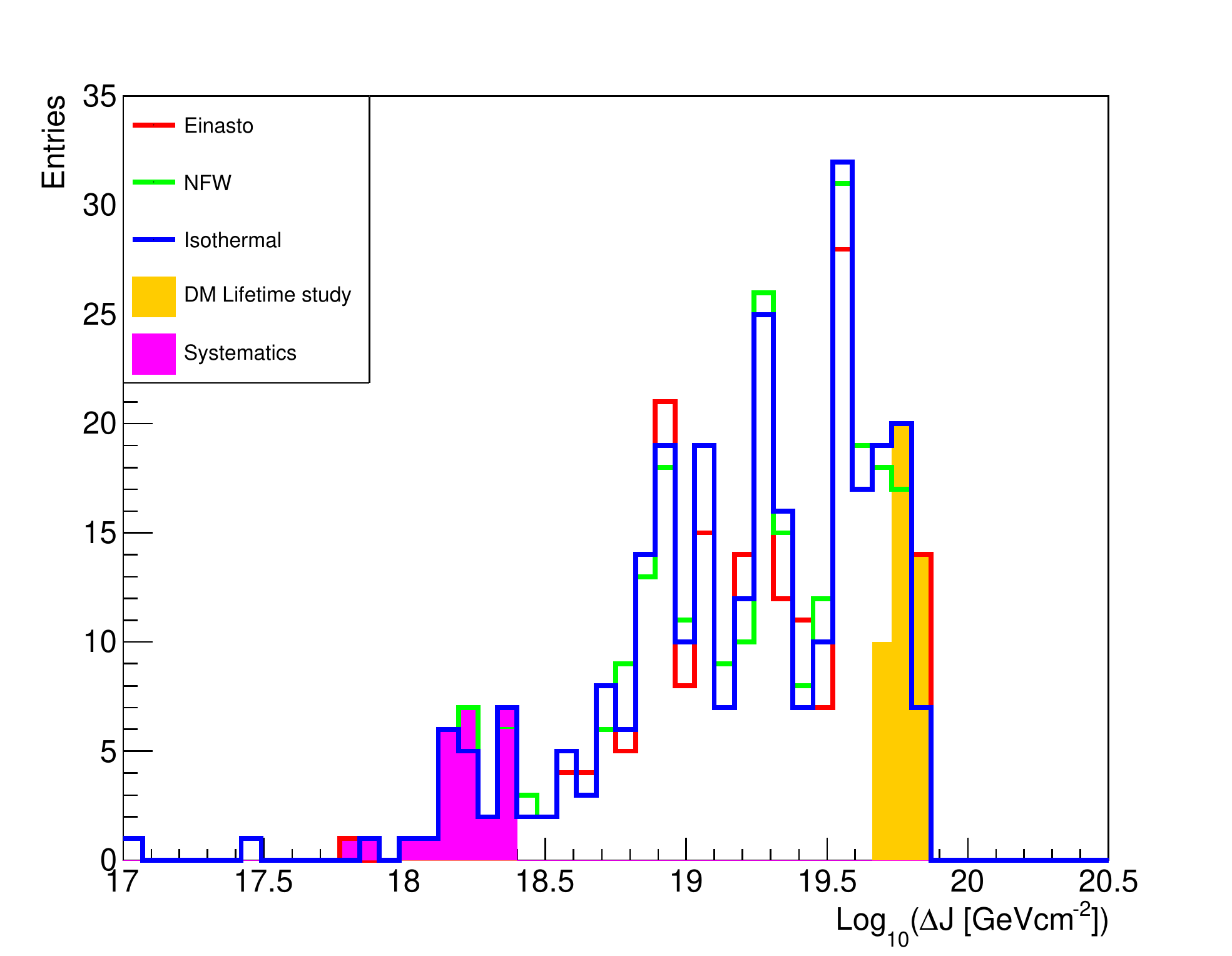}
\caption{Left: $\mathrm{Log_{10}}(\Delta J_{dec})$ computed for all the available nights of 2018, tracking one of the typical FoV used for this work. Right: distribution of the $\mathrm{log_{10}}\Delta J_{dec}$ from the left panel. The orange and magenta histograms represent the J-factor values for the selected observation nights used for the DM lifetime study and systematic evaluation, respectively. For the systematic study we did not consider the lowest J-factor available ($10^{16}\; \mathrm{GeV/cm^{-2}}$), because the time separation between ON and OFF slots was not the typical one used during the nights dedicated to the lifetime study.}
\label{fig:jfyear}
\end{figure}

We compute the OFF/ON normalization factors from the comparison of the ON and OFF number of events with high hadronness values~\footnote{The hadronness represent the output of a test statistic for particle classification (hadrons or $\gamma$-rays) computed by a Random Forest~\cite{hadronness}.}, i.e. those not passing the signal selection cut. 
In order to minimize the systematic uncertainty on the ON/OFF relative acceptance ($\sigma_{syst}$) introduced by
this procedure, we constrained our data-taking in order to observe ON and OFF under conditions as similar as possible: for each ON observation, the corresponding OFF observation was performed during the same night, only with excellent weather conditions, and following exactly the same (Zd,~Az) path in the sky. With this restriction, we have scanned every available observation night during the year 2018 (see Fig.~\ref{fig:jfyear}), and looked for 1-hour duration ON/OFF pairs with largest possible $\Delta J$ values, while keeping Zd < 35$^\circ$, $|b| > 10^\circ$ and optimizing each pointing pairs by avoiding stars with magnitude up to 6 and known $\gamma$-ray sources~\cite{3FHL} inside the camera trigger region. The $\Delta J$ has been computed starting from the curves shown in Fig.~\ref{fig:myjfactor}. We stress that in the region of the sky we are observing ($\varphi>10^\circ$), the J-factors from the three different models considered agree on the GH DM content. Thus our results will be robust against the different modelling scenario. 
We present here the results for the analysis of 20 hours of GH observation taken during 2018. 10 hours (with the lowest $\Delta J$ available) have been dedicated for evaluating $\sigma_{syst}$ since the method is highly non-standard for MAGIC (see Sec.~\ref{sec:system}), while 10 hours (with the highest $\Delta J$ available) have been used for the DM lifetime analysis (see Sec.~\ref{sec:lifetime}).
\section{Systematic errors evaluation} \label{sec:system}
\begin{figure}[t]
\centering
\includegraphics[width=0.50\textwidth]{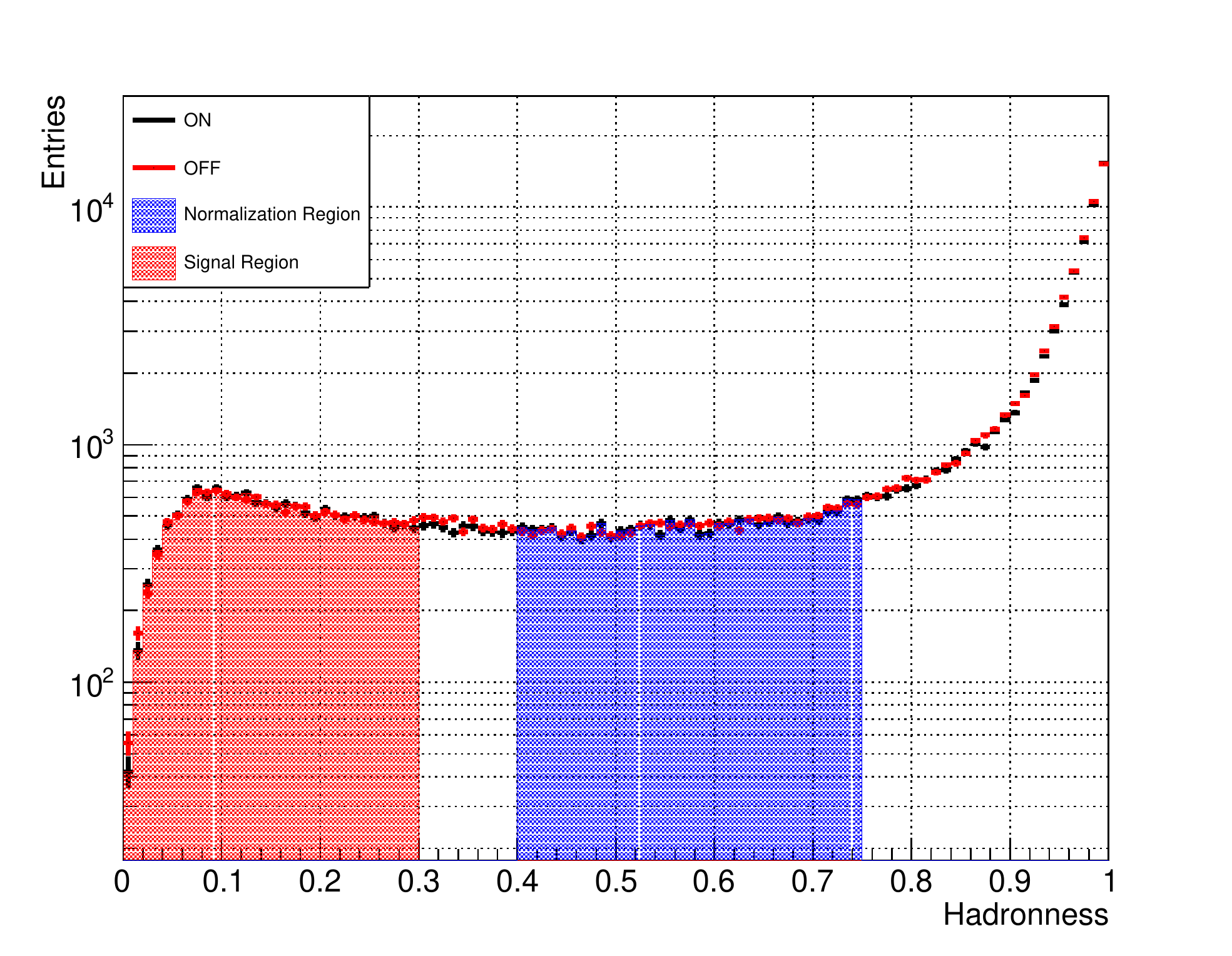}
\caption{Hadronness curves for ON and OFF data: for each night $i$, the normalization factor $\kappa_i$ is computed as the ratio of the hadronness curves in the blue region, while $R_{i}$ is computed as the normalized residual from the red region of the curves.}
\label{fig:method}
\end{figure}
\begin{figure}[t]
\centering
\includegraphics[width=0.47\textwidth]{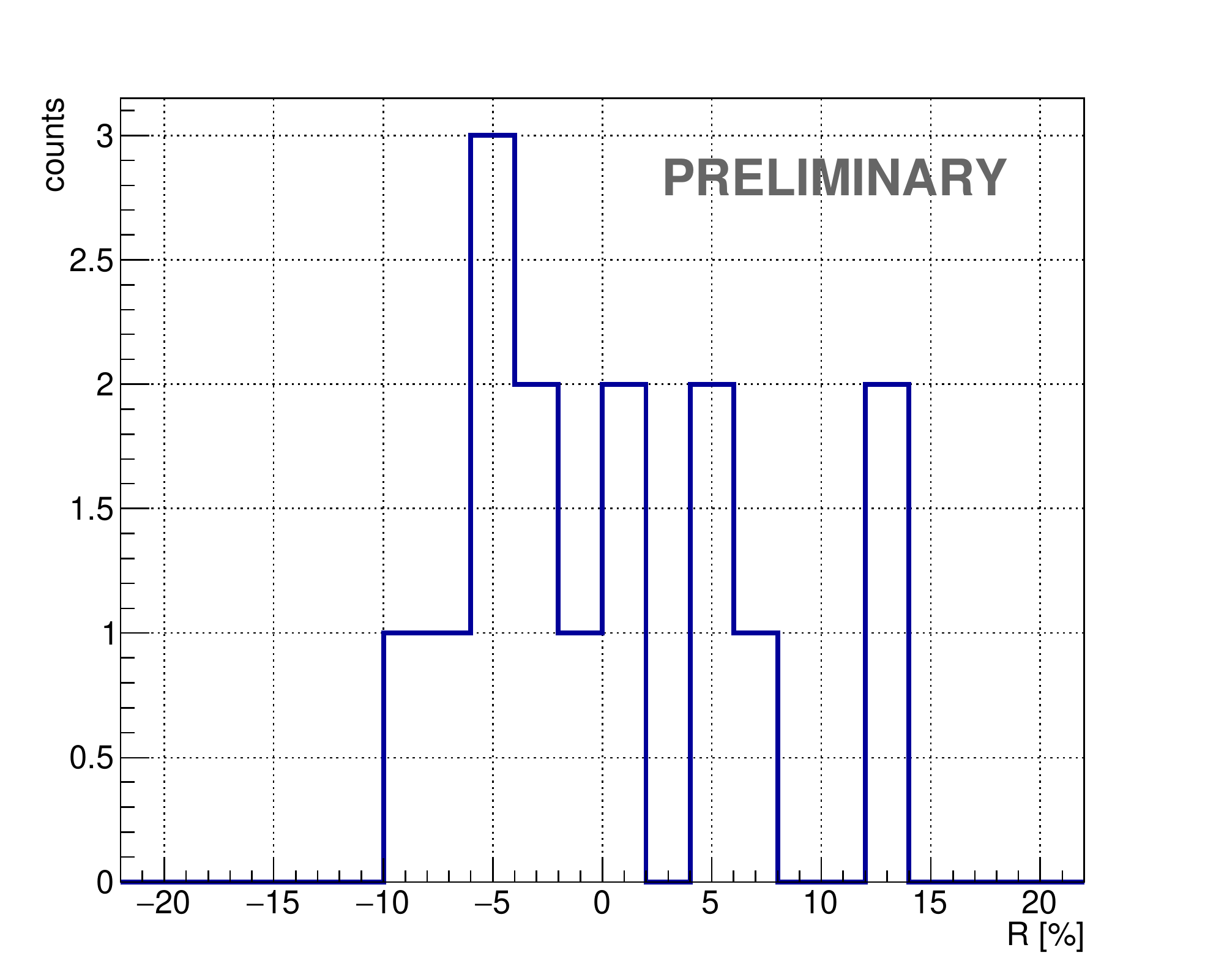}
\includegraphics[width=0.47\textwidth]{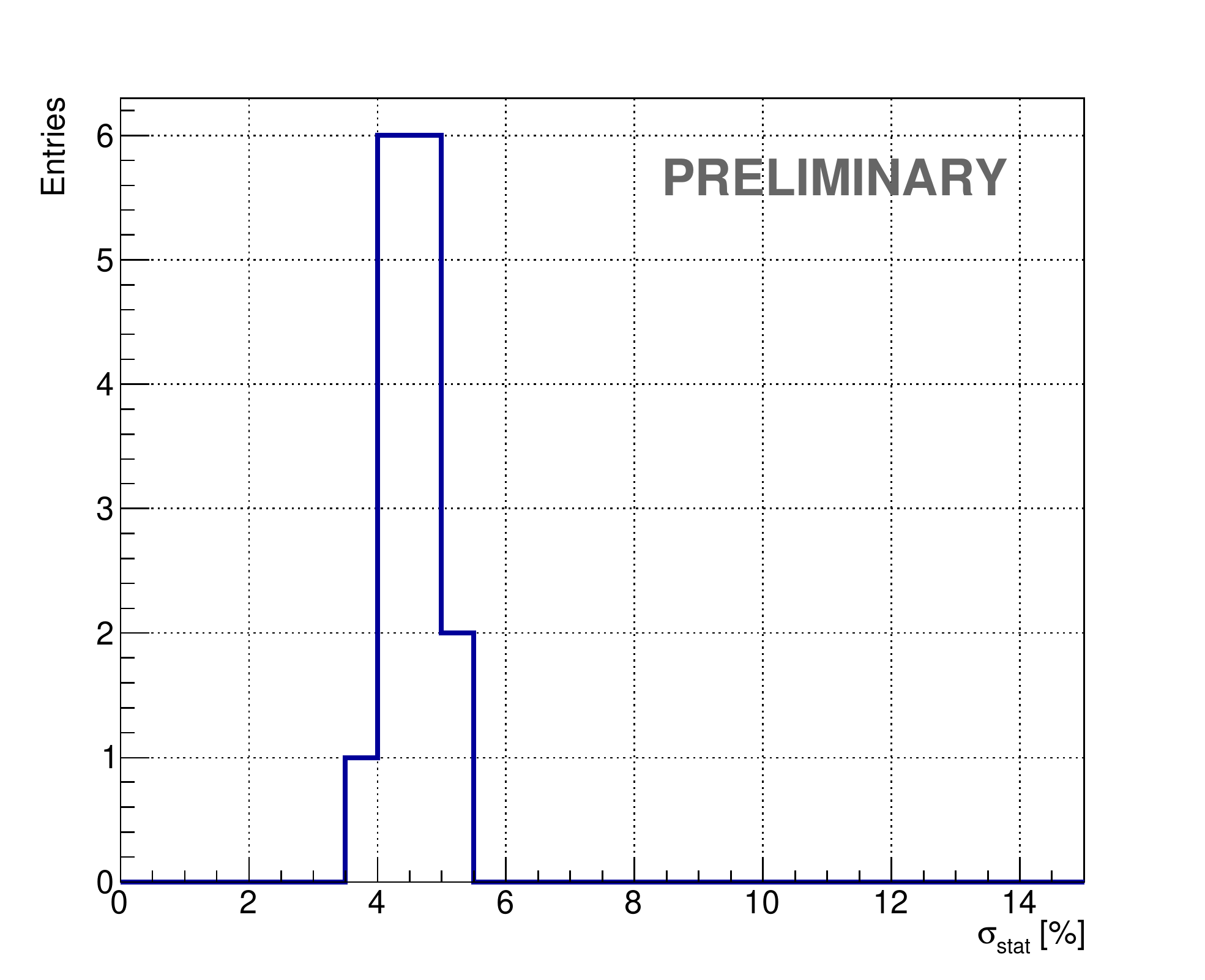}
\caption{Distribution of $R$ (left) and its statistical error $\sigma_{stat}$ (right). We compute $\sigma_{tot}$ as the sample variance of the left plot, while $\sigma_{stat}$ is the sample mean of the right one. }
\label{fig:stat}
\end{figure}
\begin{figure}[t]
\centering
\includegraphics[width=0.75\textwidth]{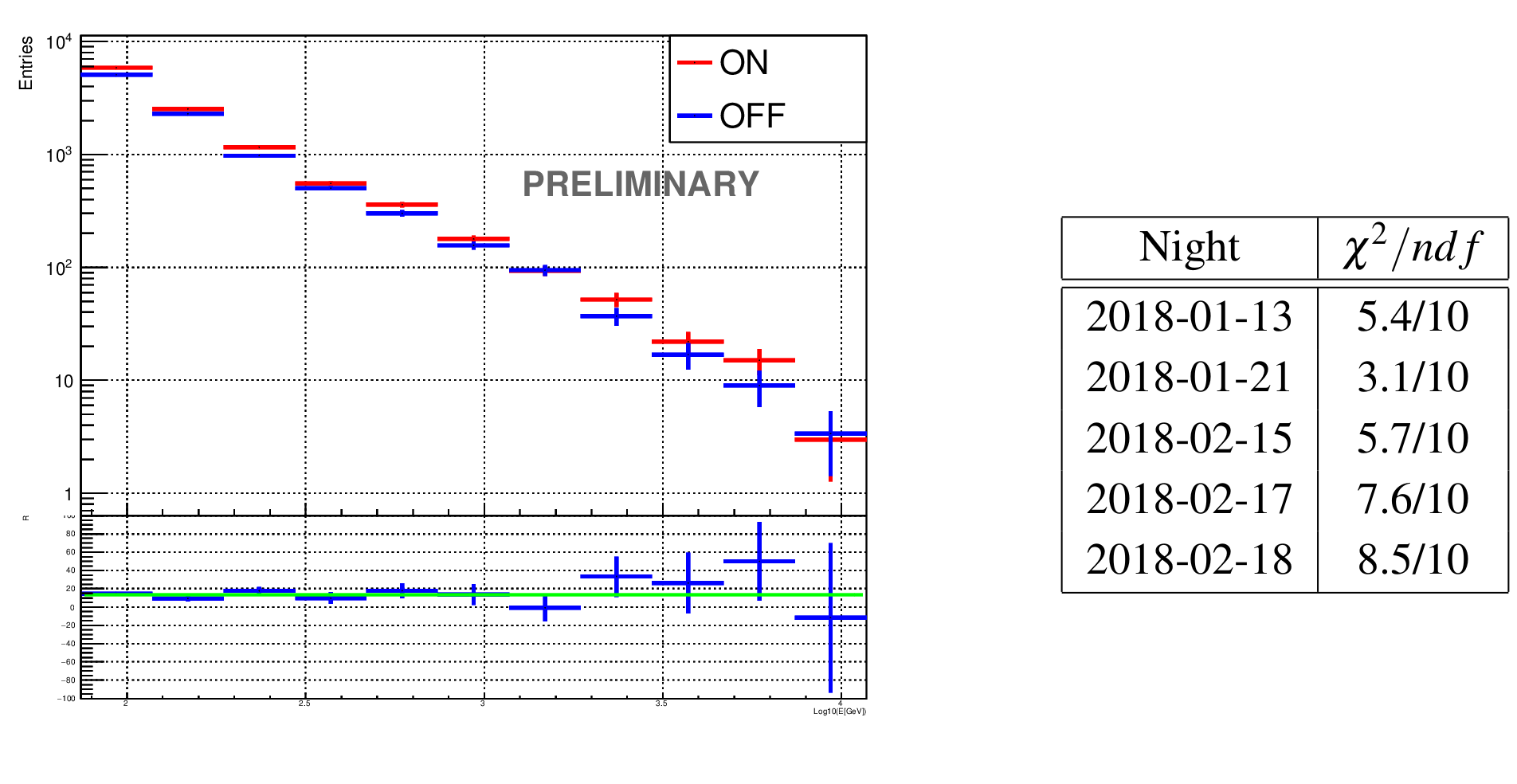}
\caption{Left: typical $\frac{dN}{dE}$ as a function of $\mathrm{log_{10}(E[GeV])}$ and the normalized residuals $R$ for 11 logarithmic equidistant energy bins are shown. The OFF curve is normalized by the same normalization factor $\kappa$ used in the analysis. Right: the table reports the $\chi^2/ndf$ values of the least-square fit of a constant line to the residual of the ON-OFF energy distribution for each night used for the evaluation of $\sigma_{syst}$. No energy dependence of the systematic uncertainties is found.} 
\label{fig:energy}    
\end{figure}

We dedicated 10 hours of GH observation for the minimization and evaluation of $\sigma_{syst}$, which is due to unknown or not controlled effects affecting the $\gamma$-ray candidate acceptance during these non-standard observations. The observation nights were selected with the criteria described previously, with the only difference of selecting ON/OFF pairs with the $\Delta J$ as low as possible. In this case, the selected nights have a $\Delta J_{syst} \approx$ 4\% of the average $\Delta J$ used for the DM search, as shown in Fig.~\ref{fig:jfyear}. The analysis cuts on the energy $E$ and the squared distance from the center of the camera $\theta^2$ were optimized by minimizing the width of the distribution of the quantity $R \equiv 2 \cdot \frac{N_{ON}-N_{OFF}/\kappa}{N_{ON}+N_{OFF}/\kappa}\%$, where $N_{ON}$ and $N_{OFF}$ are respectively the number of ON and OFF events passing all the selection cuts and $\kappa$ is the OFF/ON
normalization factor, evaluated by the ratio of events with hadronness in the interval [0.4, 0.75] (see Fig.~\ref{fig:method}). The cuts were evaluated on one half of the total data set (the training sample), subdivided into subsamples of $\sim10^\circ$ azimuth (Az) bin and the best values found are $E~>~60~\mathrm{GeV}\mathrm{\;and\;}\;\theta^2<1.44\;\mathrm{deg^2}$. These cuts were then subsequently applied to the complementary half of the dataset (the test sample) to evaluate $\sigma_{syst}$ from the corresponding distribution of $R$ (Fig.~\ref{fig:stat} left), that is expected to distribute with a total variance ~$\sigma_{tot}^{2}=\sigma_{stat}^2+\sigma_{syst}^2$, where $\sigma_{stat}=4.6\pm0.09\%$ is the sample mean of the distribution of the statistical error in the determination of $R$  (Fig.~\ref{fig:stat} right), almost constant for each Az bin subsets, since they have approximately the same observation time and conditions.
We estimated $\sigma_{tot}$ using the sample variance ($s = \frac{1}{N-1}\sum_{i=0}^N (R_{i}-\bar{R})^2$, where $N$ is the number of observed nights), and obtained $\sigma_{tot} = 6.6\pm0.75\%$. Knowing $\sigma_{tot}$ and $\sigma_{stat}$ we deduced $\sigma_{syst} = 4.8\pm1.0\%$. 
We also checked a possible dependence of
$\sigma_{syst}$ with the energy. As shown in Fig.~\ref{fig:energy}, the least-square fit of a constant value to the obtained residuals of each energy bin are compatible with the hypothesis that $\sigma_{syst}$ is
energy-independent. Thus, $\sigma_{syst}$ is introduced in our DM-search full likelihood analysis as an extra contribution (in addition to the statistical one) to the width of the Gaussian likelihood term parametrizing the uncertainty on the (energy-independent) OFF/ON normalization factor $\kappa$ (see Eq.~\ref{eq:like}). We have computed an expected worsening in the sensitivity, with respect to the $\sigma_{syst}$ = 0 case,
of a factor 1.8 for masses up to 1 TeV and 1.2 for masses up to 10 TeV.

We have cross-checked the $\sigma_{syst}$ evaluation results using MAGIC archival data that fulfil the constraints requested for GH observations: clear sky, no stars and no known VHE sources inside the FoV, low $\Delta J$ and same (Zd,~Az) path tracked for ON and OFF regions. The last request is feasible only allowing for observations during consecutive nights. This is a conservative choice, since observational conditions are more likely to be different between consecutive nights than within the same night. Thus, we selected 20+20 hours from two different extragalactic sources, observed during the GH data-taking period. 
For each of the source we extract the ON/OFF slots from the same FoV, implying by construction that  $\Delta J \approx 0$. Applying the same quality cuts, we found values for $\sigma_{syst}=6.4\pm1.0\%$ and $\sigma_{syst}=4.01\pm1.8\%$, compatible with $\sigma_{syst}$ computed in the GH case, and compatible with the no energy dependence hypothesis.
\section{Lifetime study results} \label{sec:lifetime}
We used the 10 hours acquired with the highest $\Delta J$ to perform the DM decay signal search analysis. We applied the same
analysis cuts and the same OFF normalization procedure described in the previous section. The DM analysis has been performed with the full likelihood approach described in~\cite{likelihood}, assuming no energy dependence on $\sigma_{syst}$ and no uncertainties on $\Delta J$. 
The likelihood formula is:
\begin{multline}\label{eq:like}
\mathcal{L}(1/\tau_{DM}; \nu|\mathcal{D}) =  \mathcal{J}(\Delta J|\Delta J_{obs}, \sigma_{\Delta J})\times\prod_{i=0}^{N_{samples}}\mathcal{G}(\kappa_i|\kappa_{obs},
 \sigma_{\kappa,i}) \\
 \times  \prod_{j=0}^{N_{bins}}\frac{(g_{ij}(1/\tau_{DM})+b_{ij})^{N_{ON,ij}}}{N_{ON,ij}!}\cdot e^{-(g_{ij}+b_{ij})} \times \frac{(\kappa_i b_{ij})^{N_{OFF,ij}}}{N_{OFF,ij}!}\cdot e^{-\kappa_i b_{ij}}
\end{multline}
where
$\mathcal{J}$  is the likelihood for the $\Delta J$-factor ($\sigma_{\Delta J}= 0$ in this work);
$\mathcal{D}$ and $\nu$ represent the dataset and nuisance parameters respectively;
$N_{samples}$ is the number of the azimuth bins among the whole observation nights;
$N_{bins}$ is the number of bins in estimated energy;
$g_{ij},\;b_{ij},\;N_{ON,ij}$ are respectively the estimated number of signal and background events and the number of observed events in the ON region;
$N_{OFF,ij}$ is the number of observed events in the corresponding OFF bin;
$\mathcal{G}$ is the likelihood for $\kappa_i$, parametrized by a Gaussian function with mean $\kappa_{obs,i}$ and variance $\sigma_{\kappa,i}$.
\begin{figure}[t]
\centering
\includegraphics[width=0.55\textwidth]{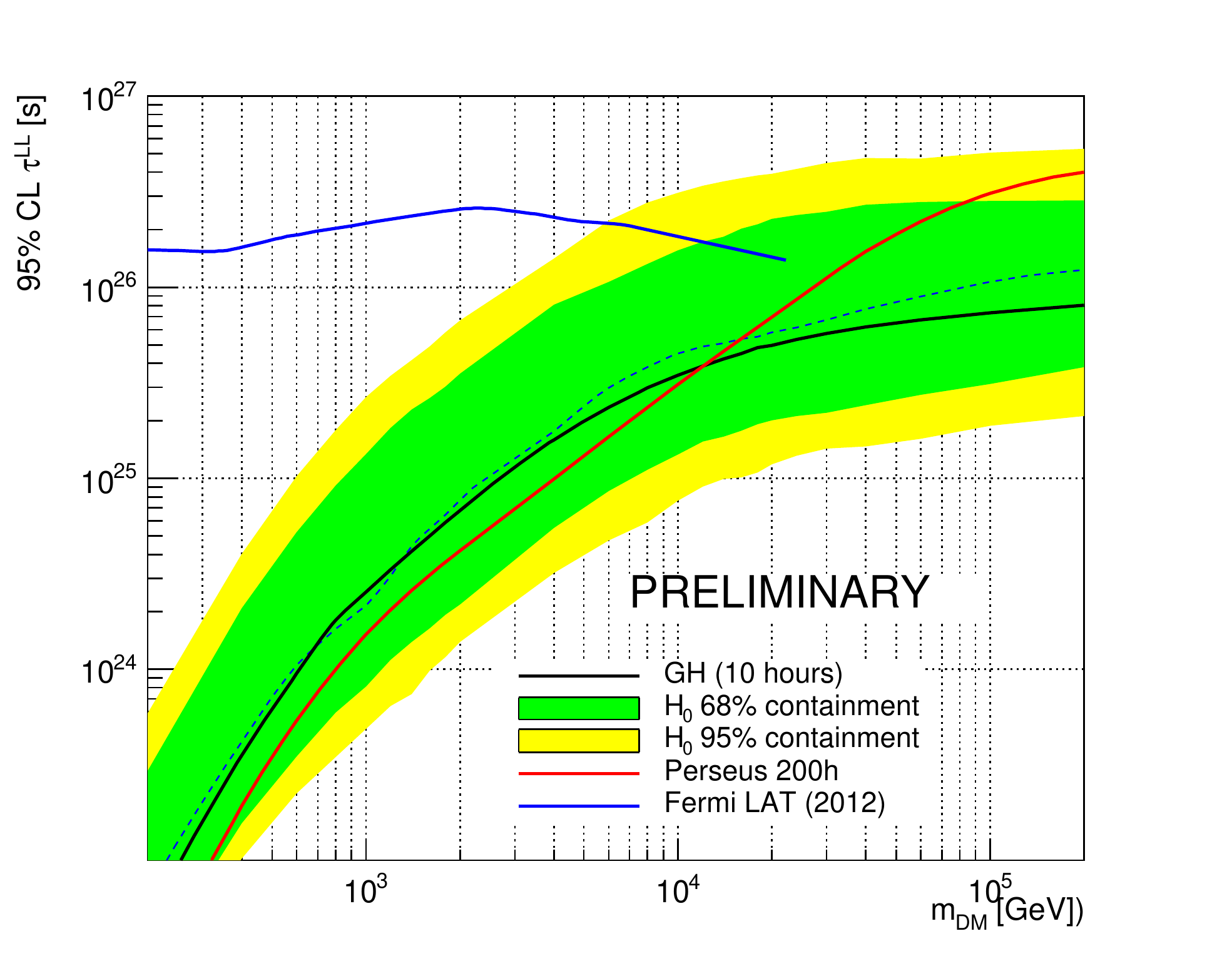}
\caption{95\% CL lower limit on the DM decay lifetime obtained with 10 hours of MAGIC GH observations, using $\sigma_{syst} = 4.8\%$ and an Einasto DM profile (solid black line), the expected limit (dashed line) and the two sided 68\% (green) and 95\% (yellow) containment bands compared to the measurements in the Perseus Galaxy Cluster by MAGIC~\cite{perseus} (red line). Limits obtained from the inner galactic center halo from the Fermi-LAT collaboration~\cite{fermi} are also shown (blue line).}
\label{fig:limit}
\end{figure}

The $\sigma_{syst}$ is taken into account in our likelihood as an additional term to the uncertainty of $\kappa$, following the formula $\sigma_{\kappa} = \sqrt{\sigma_{\kappa,stat}^2 + (\kappa\cdot\sigma_{syst})^2}$. 
We computed the 95\% CL lower limit on $\tau_{DM}^{LL}$, the life-time of DM particles decaying into $b\bar{b}$ pair (see Fig.~\ref{fig:limit}). The 10 hours of GH analysis resulted in limits as constraining as those obtained with 200 hours of observation on the Perseus Galaxy cluster~\cite{perseus} for masses up to 10 TeV, with the strongest constraint obtained for $m_{DM} = 100\;\mathrm{TeV}$ that yields to $\tau^{LL}_{DM}>10^{26}\;\mathrm{s}$. 
We show for comparison the limit obtained from the observation of the inner Galactic Halo by Fermi-LAT for $m_{DM} < 10\;\mathrm{TeV}$.

Further observations of the GH with MAGIC will allow better constraints in the future. We evaluated that increasing the total observation time to 40 hours will improve the constraints by at least a factor 2 compared to what was reported in this work.
\section{Summary}
In this contribution, we have reported the lower limits on the DM lifetime obtained with 10 hours of data
taken with the MAGIC telescopes observing the Milky Way GH. The results show the power of this method, producing one of the best limits in the literature using only a few percent of the observation time of the other DM lifetime study. Additionally we computed that the limits can be improved by a factor 2 with a limited number of additional hours of observation.
Moreover the method can be successfully applied with the future Cherenkov Telescope Array, taking the benefit of a larger FoV to increase the sensitivity to DM processes.
\section*{Acknowledgement}
MAGIC full acknowledgements: https://magic.mpp.mpg.de/acknowledgments\_ICRC2019

\end{document}